\documentclass[12pt]{article}
\usepackage{amssymb,amsmath}
\usepackage{amsfonts}
\newcommand{\be}{\begin{equation}}
\newcommand{\ee}{\end{equation}}
\newcommand{\ba}{\begin{eqnarray}}
\newcommand{\ea}{\end{eqnarray}}
\newcommand{\hook}{\raisebox{-0.35ex}{\makebox[0.6em][r]
{\scriptsize $-$}}\hspace{-0.15em}\raisebox{0.25ex}
{\makebox[0.4em][l]{\tiny $|$}}}
\begin{document}

\title{Killing forms on the five-dimensional Einstein-Sasaki 
$Y(p,q)$ spaces}

\author{Mihai Visinescu\thanks{E-mail:~~~mvisin@theory.nipne.ro}\\
{\small \it Department of Theoretical Physics}\\
{\small \it Horia Hulubei National Institute of Physics and Nuclear 
Engineering}\\
{\small \it P.O.B. MG-6,  Magurele, Bucharest, Romania}}
\date{}

\maketitle

\begin{abstract}
We present the complete set of Killing-Yano tensors on the 
five-dimensional Einstein-Sasaki $Y(p,q)$ spaces. Two new 
Killing-Yano tensors are identified, associated with the complex volume 
form of the Calabi-Yau metric cone. The corresponding hidden symmetries 
are not anomalous and the geodesic equations are superintegrable.

~

\end{abstract}

\section{Introduction}

In the last time Einstein-Sasaki geometries have become of large 
interest in connection with many modern studies in physics. In this 
paper we deal with the infinite family $Y(p,q)$ of Einstein-Sasaki 
metrics on $S^2 \times S^3$ \cite{GMSW1,GMSW2, MS,CLPP}. 
Such manifolds provide 
supersymmetric backgrounds relevant to the AdS/CFT correspondence. The 
total space $Y(p,q)$ of an $S^1$-fibration over $S^2 \times S^2$ with 
relative prime winding numbers $p$ and $q$ is topologically $S^2 \times 
S^3$.

In the present paper it will be shown that the equations of the 
geodesic motions on the $Y(p,q)$ spaces are superintegrable. For this 
purpose we present the complete set of Killing-Yano tensors which play 
an essential role for the integrability of the equations of motion.

In general a system could possess explicit and hidden symmetries 
encoded in the multitude of Killing vectors and higher rank Killing 
tensors respectively. The customary conserved quantities originate from 
symmetries of the configuration space of the system. They are 
represented by isometries of the metric generated by Killing vector 
fields. An extension of the Killing vector fields is given by the 
conformal Killing vector fields with flows preserving a given conformal 
class of metrics \cite{YAN}.

A natural generalization of Killing vector fields is represented by 
conformal Killing-Yano tensors. A conformal Killing-Yano tensor 
of rank $p$ on a $n$-dimensional Riemannian manifold $(M,g)$ 
is a $p$-form $\omega$ which satisfies:
\begin{gather}\label{CKY}
\nabla_X\omega=\frac{1}{p+1}X \hook
d\omega-\frac{1}{n-p+1}X^*\wedge d^*\omega  \,,
\end{gather}
for any vector field $X$ on $M$. Here  $\nabla$ is the Levi-Civita
connection of $g$, $X^*$ is the 1-form
dual to the vector field $X$ with respect to the metric $g$,
$\hook$ is the operator dual to the wedge product and $d^*$ is
the adjoint of the exterior derivative $d$. If $\omega$ is co-closed
in \eqref{CKY}, then we obtain the definition of a Killing-Yano
tensor \cite{YAN}. Killing-Yano tensors are also called Yano tensors 
or Killing forms. A particular class of Killing forms is represented 
by the special Killing forms which satisfy, for some constant $c$, 
the additional equation \cite{US}:
\be\label{SKY}
\nabla_X(d\omega)=c \, X^* \wedge\omega \,,
\ee
for any vector field $X$ on $M$. Let us note that most known Killing 
forms are special. 

Besides the antisymmetric generalizations of the Killing vectors one 
might also consider higher order symmetric tensors. A covariant 
symmetric field $K$ of rank $r$ on a Riemannian manifold $(M,g)$ is a 
St\"ackel-Killing tensor field if the symmetrization of its covariant 
derivative vanishes identically
\[
\nabla_{(\lambda}K_{\mu_1,\dots ,\mu_r)} = 0\,.
\]
These symmetric tensors are associated with conserved quantities of 
degree $r$ in the momentum variables, and generate symplectic 
transformations in the phase space of the system \cite{MC}.

These two generalizations of the Killing vectors could be related. 
Given two rank $r$ Killing forms $\omega_{\mu_1 \dots \mu_r}$ and
 $\sigma_{\mu_1 \dots \mu_r}$ it is possible to associate with them a 
St\"ackel-Killing tensor of rank $2$
\be\label{KYY}
K^{(\omega,\sigma)}_{\mu\nu} = ( \omega_{\mu\lambda_2 \dots \lambda_r}
\sigma_{\nu}^{\phantom{\nu}\lambda_2 \dots \lambda_r} +
\sigma_{\mu\lambda_2 \dots \lambda_r}
\omega_{\nu}^{\phantom{\nu}\lambda_2 \dots \lambda_r})\,.
\ee

At the quantum level the operator $\nabla_\mu K^{\mu\nu} \nabla_\nu$ 
may be thought of as the equivalent of the classical conserved quantity 
$K_{\mu\nu}\dot{\gamma}^\mu \dot{\gamma}^\nu$ which is constant along 
geodesics $\gamma$.
The remarkable fact is that  $\nabla_\mu K^{\mu\nu} \nabla_\nu$ 
operating on scalar fields commutes with the Klein-Gordon operator 
$\square = \nabla_\mu g^{\mu\nu} \nabla_\nu$ \cite{BC}. 
Therefore, when the 
St\"ackel-Killing tensor $K_{\mu\nu}$ is of the form \eqref{KYY}, there 
are no quantum anomalies thanks to an integrability condition satisfied 
by the Killing-Yano tensors \cite{MC,BC,EPS}.

The organization of this paper is as follows: In the next Section we 
present what is essentially a brief review of how Einstein-Sasaki 
manifolds can be constructed as $U(1)$ bundles over Einstein-K\"ahler 
manifolds. Further an Einstein-Sasaki metric may be defined as a 
$(2n+1)$-dimensional Einstein metric such that the cone over it is a 
K\"ahler, Ricci-flat metric of complex dimension $n+1$. In Section 3 we 
restrict to the five-dimensional $Y(p,q)$ manifolds and present the 
complete set of Killing forms. Finally we give our conclusions in 
Section 4.

\section{Progression from Einstein-K\"ahler to \\ Einstein-Sasaki to 
Calabi-Yau manifolds}

A Riemannian manifold $(M_{2n+1},g_{S})$ of odd dimension $2n+1$ is 
Sasakian if and only if its metric cone 
\[
C(M_{2n+1}) = \mathbb{R}_{>0} \times M_{2n+1} , \quad 
g_{\text{cone}} = dr^2 + r^2 g_{S}\,,
\]
is K\"ahler with the complex dimension $n+1$.
Moreover a Sasakian metric $g_{ES}$ is Einstein with 
$\text{Ric} = 2n\, g_{ES}$ 
if and only if its metric cone is Ricci-flat and K\"ahler, i. e. 
a Calabi-Yau manifold \cite{SP}.

On the other hand Sasakian manifolds can be constructed as principal 
$S^1$-bundle over a K\"ahler manifold $M_{2n}$. It is convenient to 
write the Einstein-Sasaki metric in the form \cite{GHP,DK}
\[
g_{ES} = g_{EK} + (d\alpha + \sigma)^2\,,
\]
where $g_{EK}$ is the metric of the Einstein-K\"ahler manifold $M_{2n}$ 
with the complex dimension $n$ and 
\be\label{Eta}
\eta = d\alpha + \sigma\,,
\ee
is the Sasakian $1$-form of the Einstein-Sasaki metric.
The form $\sigma$ satisfies
\be\label{JEK}
d\sigma = 2 \Omega_{EK}\,,
\ee
where $\Omega_{EK}$ is K\"ahler form of the Einstein-K\"ahler base 
manifold $M_{2n}$. In consequence the K\"ahler form of the Calabi-Yau 
cone manifold can be written as
\be\label{omegacone}
\Omega_{cone} = r dr\wedge (d\alpha + \sigma) + r^2 \Omega_{EK}\,.
\ee

In local holomorphic coordinates $(z^1,...,z^m)$, the
associated K\"{a}hler form $\Omega$ of a K\"ahler manifold $M_m$ of 
complex dimension $m$ is
\[
\Omega=i g_{j\bar{k}}dz^j\wedge d\bar{z}^k \,,
\]
and the volume form (which is just the Riemannian volume form 
determined by the metric) is a real $(m,m)$-form
\[
d\mathcal{V}=\frac{1}{m!}\Omega^m\,,
\]
where  $\Omega^m$ is the wedge product of $\Omega$ with itself $m$ 
times. If the volume of a K\"{a}hler manifold is written as
\[
d\mathcal{V} = dV \wedge d{\bar{V}}\,,
\]
then $dV$ is the complex volume holomorphic $(m,0)$-form of $M$ 
\cite{YO,MVGV}. 

There is a 1-1-correspondence between the special Killing $p$-forms 
\eqref{SKY} on a compact oriented simply connected Riemannian manifold 
$M$ and parallel $(p+1)$-forms on the cone manifold $C(M)$. Taking into 
account that the metric cone $C(M)$ is either flat or irreducible, the 
existence of Killing forms on the manifold $M$ can be settled 
investigating the parallel forms on flat or irreducible manifolds. 
These forms are classified by means of holonomy groups \cite{MB,BES}.
In the case of a Sasakian base manifold $M_{2n+1}$ the following two 
possibilities are of interest \cite{US}:
\begin{enumerate}
\item{ The metric cone $C(M_{2n+1})$ has holonomy $U(n+1)$ and 
equivalently is K\"ahler, $M_{2n+1}$ is Sasaki and all special Killing 
forms are described by the $(2k+1)-$forms:
\be\label{omega}
\Psi_k = \eta \wedge (d\eta)^k \quad , \quad k = 0, 1,
\cdots , n-1\,.
\ee
Besides these Killing forms, there are $n$ closed conformal Killing
forms (also called $\ast$-Killing forms)
\[
\Phi_k = (d\eta)^k \quad , \quad k = 1, \cdots , n \,.
\]}
\item{ The holonomy of $C(M_{2n+1})$ is $SU(n+1)$, the metric cone is 
K\"ahler and in addition Ricci flat, or equivalently $M_{2n+1}$ is 
Einstein-Sasaki. In this situation there are {\it two additional} 
Killing forms of degree $n$ on the manifold
$M_{2n+1}$. These additional Killing forms are connected with the 
additional parallel forms of the Calabi-Yau cone manifold $M_{2n+2}$ 
given by the complex volume form and its conjugate \cite{US} which can 
also be described using the Killing spinors of an Einstein-Sasaki 
manifold \cite{CB}.}
\end{enumerate}

\section{$Y(p,q)$ manifolds}

The starting point is the explicit local metric of the 5-dimensional 
$Y(p,q)$ manifold given by the line element \cite{GMSW1,GMSW2,BK,RS}
\begin{align}\label{Ypq}
ds^2_{ES} & = \frac{1-c\, y}{6}( d \theta^2 + \sin^2 \theta\, d \phi^2) 
+  \frac{1}{w(y)q(y)} dy^2 
+ \frac{q(y)}{9} ( d\psi - \cos \theta \, d \phi)^2 \nonumber\\
& \quad  + 
w(y)\left[ d\alpha + \frac{ac -2y+ c\, y^2}{6(a-y^2)}
(d\psi - \cos\theta \, d\phi)\right]^2\,,
\end{align}
where
\[
w(y)  = \frac{2(a-y^2)}{1-cy} \,, \quad 
q(y)  = \frac{a-3y^2 + 2c y^3}{a-y^2}\,,
\]
and $a,c$ are constants. A detailed discussion of the range of these 
parameters is given in \cite{GMSW2} in connection with the regularity 
properties of the $Y(p,q)$ metrics. For $c=0$ the metric takes the 
local form of the standard homogeneous metric on $T^{1,1}$ 
\cite{MS2006}. Otherwise the constant $c$ can be rescaled by a 
diffeomorphism and in what follows we assume $c=1$.

The coordinate change $\alpha = -\frac{1}{6}\beta - \frac{1}{6} \psi'
\,,\, \psi=\psi' $
takes the line element \eqref{Ypq} to the following form
\[
\begin{split}
ds^2_{ES} & = \frac{1- y}{6}( d \theta^2 + \sin^2 \theta\, d \phi^2) 
+  \frac{1}{p(y)} dy^2 
+ \frac{1}{36} p(y) ( d\beta +  \cos \theta \, d \phi)^2 \nonumber\\
& \quad + 
\frac{1}{9}[d\psi' - \cos\theta \, d\phi +
y ( d\beta +  \cos \theta \, d \phi)]^2 \,,
\end{split}
\]
with $p(y) = w(y) \, q(y)$.

The Sasakian $1$-form of the $Y(p,q)$ space is
\be\label{eta}
\eta = \frac{1}{3} d \psi' + \sigma \,,
\ee
with
\[
\sigma = \frac{1}{3}[- \cos\theta \, d\phi +
y ( d\beta + \cos \theta \, d \phi)]\,,
\]
connected with local K\"ahler form $\Omega_{EK}$ as in \eqref{JEK}.
Note also that the holomorphic $(2,0)$-form of the Einstein-K\"ahler 
base manifold $M_4$ is
\[
dV_{EK} = \sqrt{\frac{1-y}{6p(y)}}(d\theta + i \sin\theta\,
d\phi) 
\wedge \Bigl[ dy + i\frac{p(y)}{6} (d\beta + \cos\theta \, d\phi)
\Bigr]\,.
\]

The form of the metric \eqref{Ypq} with the $1$-form \eqref{eta} is the 
standard one for a locally Einstein-Sasaki metric with 
$\frac{\partial}{\partial \psi'}$ the Reeb vector field.

From the isometries $SU(2) \times U(1) \times U(1)$ the momenta 
$P_\phi, P_\psi, P_\alpha$ and the Hamiltonian describing the geodesic 
motions are conserved \cite{BK,RS}. $P_\phi$ is the third component of 
the $SU(2)$ angular momentum, while $P_\psi$ and  $P_\alpha$ are 
associated with the $U(1)$ factors. Additionally, the total $SU(2)$ 
angular momentum given by
\be\label{J2}
J^2 = P^2_\theta + \frac {1}{\sin^2\theta}(P_\phi + \cos\theta 
P_\psi)^2 + P^2_\Psi\,,
\ee
is also conserved.

In what follows we are looking for further conserved quantities 
specific for motions in Einstein-Sasaki spaces. First of all, on the 
Sasakian manifold with the $1$-form \eqref{Eta}, the form \eqref{omega}
for $k=1$
\be\label{psi}
\begin{split}
\Psi & = \frac{1}{9} [(1-  y)\sin\theta \, d\theta \wedge d\phi \wedge 
d\psi' + dy \wedge d \beta \wedge d \psi' \\ 
& \quad +  \cos\theta \,
dy\wedge d\phi\wedge d\psi' - \cos\theta \, dy \wedge d\beta \wedge 
d\phi \\ 
& \quad + (1-  y) y 
\sin\theta \, d\beta \wedge d\theta \wedge d\phi]\,,
\end{split}
\ee
is a special Killing form \eqref{SKY}. Let us note also that 
\[
\Psi_k = (d\eta)^k \,, \quad k= 1, 2\,,
\]
are closed conformal Killing forms \cite{DK,MVGV}.

On the Calabi-Yau manifold $C(M_{2n+2})$ the K\"ahler form 
\eqref{omegacone} with the Sasakian $1$-form \eqref{eta} is
\[
\begin{split}
\Omega_{cone} & = r^2 \frac{1 - y}{6}\sin\theta\, d\theta \wedge d\phi + 
\frac{r^2}{6} d y \wedge (d\beta + \cos\theta\, d\phi)\\
& \quad  + \frac{1}{3}r dr\wedge 
[y\, d\beta + d\psi' - (1 - y)\cos\theta\, d\phi] \,.
\end{split}
\]

The complex volume holomorphic $(3,0)$-form on the metric cone is
\cite{MS2006}
\begin{equation}\label{Vcomplex}
\begin{split}
dV_{cone} & = e^{i\psi'} r^2 d V_{EK} \wedge ( dr + i r \wedge\eta)\\
& = e^{i\psi'} r^2 \sqrt{\frac{1 - y}{6p(y)}}(d\theta + i \sin\theta\,
d\phi) \\
& \quad \wedge \Bigl[ dy + i\frac{p(y)}{6} (d\beta + \cos\theta \, d\phi)
\Bigr]\\
& \quad \wedge \Bigl[dr + i \frac{r}{3} [y\, d\beta + d\psi' - 
(1 - y)\cos\theta \, d\phi]\Bigr]\,.
\end{split}
\end{equation}

To find explicitly the additional Killing forms on the $Y(p,q)$ spaces 
we shall consider the additional parallel forms on the metric cone 
related to the complex volume form \eqref{Vcomplex} and its conjugate. 
As real forms they are given by the real and respectively imaginary 
part of the volume form.
For this purpose we make use of the fact that for any $p$-form 
$\omega^M$ on the space $M_{2n+1}$ we can define an associated 
$(p+1)$-form $\omega^C$ on the cone $C(M_{2n+1})$
\begin{equation}\label{psic} 
\omega^C := r^p dr \wedge \omega^M + \frac{r^{p+1}}{p+1} d\omega^M \,.
\end{equation}
Moreover $\omega^C$ is parallel if and only if $\omega^M$ is a special 
Killing form \eqref{SKY} with constant $c= -(p+1)$ \cite{US}.
Therefore the 1-1-correspondence between special Killing $p$-forms on 
$M_{2n+1}$ and parallel $(p+1)$-forms on the metric cone $C(M_{2n+1})$
allows us to describe the additional Killing forms on Einstein-Sasaki 
$Y(p,q)$ spaces.

Extracting from the complex volume form \eqref{Vcomplex} the form 
$\omega^M$ on the Einstein-Sasaki space  according to \eqref{psic} 
for $p=2$  we get the following additional Killing $2$-forms of the 
$Y(p,q)$ spaces written as real forms:
\be\label{xi}
\begin{split}
\Xi &= \mathrm{Re}\, \omega^M
= \sqrt{\frac{1 - y}{6\, p(y)}}\\
&\quad \times \biggl( \cos\psi'
\Bigl[- dy \wedge d\theta + \frac{p(y)}{6} \sin\theta \,  d\beta 
\wedge d\phi \Bigr] - \biggr.\\
& \quad \biggl. - \sin\psi'\Bigl[- \sin\theta \, dy \wedge d\phi
- \frac{p(y)}{6} d\beta \wedge d\theta  
+ \frac{p(y)}{6} \cos\theta \, d\theta \wedge d\phi \Bigr] \biggr)\,,
\end{split}
\ee
\be\label{up}
\begin{split}
\Upsilon &= \mathrm{Im}\, \omega^M
= \sqrt{\frac{1 - y}{6\, p(y)}}\\
&\quad \times \biggl( \cos\psi'
\Bigl[- \sin\theta \, dy \wedge d\phi
- \frac{p(y)}{6} d\beta \wedge d\theta  
+ \frac{p(y)}{6} \cos\theta \, d\theta \wedge d\phi \Bigr] - \biggr.\\
& \quad \biggl. + \sin\psi'\Bigl[- dy \wedge d\theta + 
\frac{p(y)}{6} \sin\theta \,  d\beta \wedge d\phi \Bigr]\biggr)\,,
\end{split}
\ee

The St\"ackel-Killing tensors associated with the Killing forms
$\Psi$ \eqref{psi}, $\Xi$ \eqref{xi}, $\Upsilon$ \eqref{up}
are constructed as in \eqref{KYY}. The list 
of the non vanishing components of these St\"ackel-Killing tensors is 
quite long and will be given elsewhere. 
Together with the Killing vectors $P_\phi, P_\psi, P_\alpha$ and the 
total angular momentum $J^2$ \eqref{J2} these St\"ackel-Killing tensors 
provide the superintegrability of the $Y(p,q)$ geometries.

\section{Conclusions}

In this paper we have presented the complete set of Killing forms on 
five-dimen\-sio\-nal Einstein-Sasaki $Y(p,q)$ spaces. The multitude of 
St\"ackel-Killing tensors associated with these Killing forms implies 
the superintegrability of the geodesic motions. 

In connection with the third rank Killing-Yano tensors on the $Y(p,q)$ 
spaces let us note an interesting geometrical interpretation of the Lax 
representation \cite{RG,GKR,AM}.

In the theory of the classical spinning particles additional 
non-generic supersymmetries are generated from Killing-Yano tensors. At 
the quantum level from Killing forms one can construct
Carter-McLenagham like operators \cite{CL} which commute with the 
standard Dirac operator. It is also worth noting 
that in the full quantum theory the symmetries generated by Killing 
forms are not anomalous 
\cite{MC}.

These remarkable properties of the Killing forms offer new perspectives 
in the investigation of the supersymmetries, separability of 
Hamilton-Jacobi, Klein-Gordon, Dirac equations on $Y(p,q)$ spaces.
\subsection*{Acknowledgments}
Support through CNCS-UEFISCDI project number PN-II-ID-PCE-2011-3-0137 
is acknowledged.

\end{document}